\documentclass[doublecol]{epl2} 
\usepackage{amsmath}

\usepackage{graphicx}
\usepackage{dcolumn}
\usepackage{bm}

\title{Derivation of the Disorder Induced Interaction and
the Phase Diagram of Cuprate Superconductors}
\shorttitle{The Disorder Induced Interaction of Cuprates}

\author{E. V. L. de Mello, and Raphael B. Kasal}

\institute{
  \inst{1} Instituto de F\'isica, Universidade Federal Fluminense, Niter\'oi, RJ 24210-340, Brazil
}
\date{\today}

\abstract{
We show that an electronic phase transition described by the 
Cahn-Hilliard equation has important applications to cuprate
superconductors. The simulations of the local charge density and free
energy reveal two main features: i) The segregation process
creates tiny isolated regions with potential wells where the holes can be
bound in single-particle levels. ii) The clustering process also
gives rise to an effective two-body pairing interactions and
superconducting amplitudes $\Delta_{sc}(\vec r)$ at
low temperatures.  The resulting system resembles a
granular superconductor with the resistivity transition driven by 
Josephson coupling among these nanoscale grains. This approach 
reproduces the well known critical temperature transition $T_c(p)$ as 
function of the doping level $p$.  Furthermore, the local density of states 
with spatial dependent gaps $\Delta(\vec r)$ is due to the intragrain 
single-particle bound states that remain above $T_c$, 
which characterizes the pseudogap phase and reproduces
many measurements.
}
\begin{document}

\maketitle
\section{Introduction}

The origin of the superconducting gap associated with the superconducting
state and the relation to the pseudogap above the transition
temperature ($T_c$) remains one of the central questions in high-$T_c$ 
research\cite{TS,Tallon,Zawa}. It is a matter of  debate whether this
problem is connected with the charge inhomogeneities observed
in many experiments\cite{Tranquada,Bianconi,Singer} and with 
the position dependent energy gap $\Delta(\vec r)$ measured
by Scanning Tunneling Microscopy (STM)\cite{Pan,McElroy,
Gomes,Pasupathy,Kato,Pushp,Kato2}. In this letter we
suggest that these two problems are strongly related 
and their solution must be worked out together. 

There are steadily accumulating evidences that the charge distribution 
is microscopically inhomogeneous in the $CuO_2$ planes of
high temperature superconductors (HTSC). The most well known charge
inhomogeneity is the charge stripe structure\cite{Tranquada,Bianconi}.
Nuclear quadruple resonance (NQR) have measured
two signals  corresponding to two (low and high) densities
around the average doping\cite{Singer}, that increases as 
the temperature is decreased. This evolution with temperature
can be an evidence of an electronic phase separation (EPS). More 
recently, NQR experiments
indicated that the charge inhomogeneity in the planes is
correlated with the $O$ dopant atoms in $YBa_2Cu_3O_y$ 
system\cite{Keren}. Also a granular fractal microstructure
of interstitial $O$ was recently measured in $La_2Cu_{4+y}$\cite{Bianconi2}
Similarly, recent STM data have revealed one of the
most puzzling property of cuprates, the
non-uniform energy gaps $\Delta(\vec r)$ on the length scale of
nanometers\cite{Pan,McElroy,Gomes,Pasupathy,Kato,Pushp,Kato2}. 
The derived LDOS have generally two forms: one type with smaller gaps
and well defined peaks and other with larger gaps and ill-defined
peaks\cite{McElroy}. Furthermore, some of these gaps remain well above the 
superconducting critical temperature $T_c(p)$\cite{Gomes,Pasupathy}. 
These two types of gaps\cite{Zawa} may occur due to the existence of two energy 
scales that have been observed on electronic Raman scattering 
measurements\cite{LeTacon}, STM data\cite{Kato,Pushp,Kato2}, 
Angle Resolved Photon Emission 
(ARPES)\cite{Shen} and combined STM-ARPES\cite{Mad}. However,
the origin and even the existence of this two-gap picture is
still a matter of debate\cite{Campuzano,Chatterjee}.

In order to obtain an unified interpretation to all of these 
experiments we study an 
EPS transition that generates regions of low and high densities. 
Such transition may be driven by the lower free energy of undoped
antiferromagnetic (AF) regions\cite{Mello09} (intrinsic) or by the
out of plane dopants\cite{Keren} (extrinsic origin) like
oxygen interstitials\cite{Bianconi2}.
As $p$ increases, the Coulomb repulsion among the charges
in the local high doping  regions increases
the energy cost of the phase separation, which ceases the 
transition in the overdoped region, in agreement with the
disappearance of the local AF fluctuations\cite{Tranquada2}.
This clustering phenomenon gives rise to a potential that
bunch up the holes and separates low and high density phases in
the form of small domains or grains in the $CuO_2$ planes. 
This EPS process can be regarded as due to an effective
attractive potential, which is used in the self-consistent
calculations to obtain the intragrain superconducting amplitudes. 
In this approach the resistivity 
transition temperature $T_c(p)$ occurs when the Josephson energy 
$E_J(p)$ among the separated regions is equal to $K_BT_c(p)$.

\section{The Potential from the Charge Inhomogeneities}
\begin{figure}[ht]
\begin{center}

     \centerline{\includegraphics[width=7.0cm]{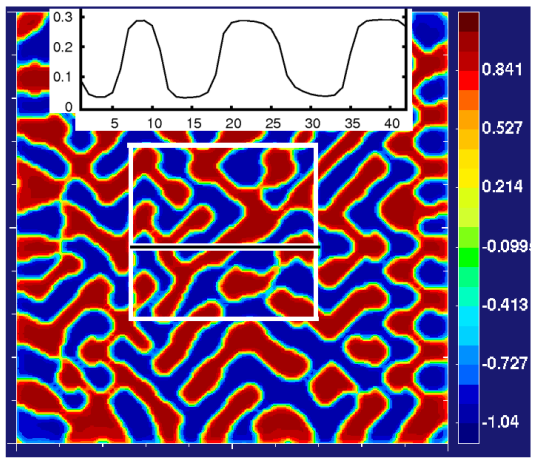}}
\caption{ (color online) The density map simulation of the 
inhomogeneous charge density on a grid with $105 \times 105$
sites. The white square indicates where the superconducting 
calculations are made. For the $p=0.16$ compound, the local densities 
$p_i$ along the line with a black trace are shown on the top of the figure. 
}
\label{MapVu} 
\end{center}
\end{figure}

\begin{figure}[ht]
\begin{center}

     \centerline{\includegraphics[width=7.0cm]{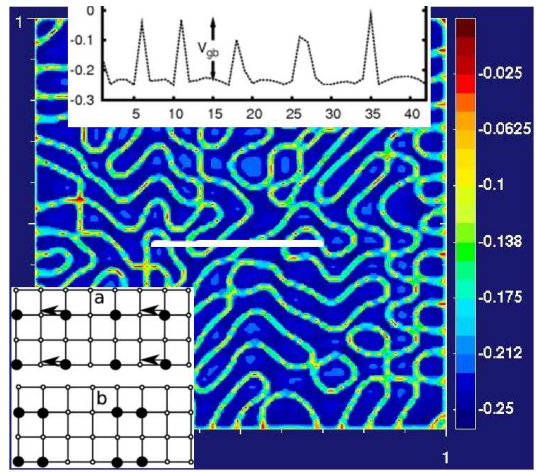}}
\caption{ (color online) The simulation of the potential ${\it V}(p,i,T)$. 
The  values on 42 sites at the white line
are shown in the top inset to demonstrate the potential wells with
average barrier $\approx V_{gb}$ where
single-particle bound states are formed. The low inset shows schematically 
an example of the ${\it V}(p,i,T)$ minimization with the holes (black dots) 
clustering by the CH diffusion {\it as an effective two-body hole attraction}.
}
\label{EV6200} 
\end{center}
\end{figure}

Phase separation is a very general
phenomenon in which a structurally and chemically homogeneous
system shows instability toward a disordered composition\cite{Bray}.
As discussed above, it has been recognized to be very important to cuprate 
superconductors\cite{Muller} but there is not a theory to
describe the degree of disorder as function of doping and temperature. 
Consequently, to deal with this problem, we use 
the general theory of Cahn-Hilliard (CH)\cite{CH,Otton,Mello04} 
adapted to the cuprates. The starting point is the EPS temperature
$T_{PS}(p)$, taken close to the upper pseudogap that produces
an anomaly measured in some experiments\cite{TS,Tallon,Zawa}. 
The transition order parameter is the difference 
between the local and the average charge density $u(p,i,T)\equiv (p(i,T)-p)/p$.
Clearly $u(p,i,T)=0$ corresponds to the homogeneous system
above $T_{PS}(p)$. Then the typical Ginzburg-Landau free energy
functional in terms of $u(p,i,T)$  is
given by

\begin{eqnarray}
f(u)= {{{1\over2}\varepsilon^2 |\nabla u|^2 +V(u,T)}}.
\label{FE}
\end{eqnarray}
Where the potential ${\it V}(u,T)= -A^2(T)u^2/2+B^2u^4/4+...$,
$A^2(T)=\alpha(T_{PS}(p)-T)$, $\alpha$ and $B$ are constants. 
$\varepsilon$ gives the
size of the boundaries between the low and high density phases
\cite{Otton,Mello04}. The CH equation can be written\cite{Bray} 
in the form of a continuity equation of the local density of free energy $f$,
$\partial_tu=-{\bf \nabla.J}$, with the current ${\bf J}=M{\bf
\nabla}(\delta f/ \delta u)$, where $M$ is the mobility or the
charge transport coefficient that sets the time scale. Therefore,
\begin{eqnarray}
\frac{\partial u}{\partial t} = -M\nabla^2(\varepsilon^2\nabla^2u
- A^2(T)u+B^2u^3).
\label{CH}
\end{eqnarray}
In Fig.(\ref{MapVu}), we show a typical simulation of the density map  
with the two (hole-rich and hole-poor) phases given by
different colors.

The potential ${\it V}(p,i,T)$
isolates the hole-rich and hole-poor regions forming  grain
boundaries between these two phases\cite{Mello04}.  
${\it V}(p,i,T)$ reaches its minimum value at the low
and high equilibrium densities as it
is demonstrated by the simulations shown in Fig.(\ref{MapVu})
and Fig.(\ref{EV6200}).
Our most important finding is that ${\it V}(p,i,T)$ produces
two different effects: first, it creates  potential wells
as it is shown in the top inset by the values of 
${\it V}(p,i,T)$ on 42 sites along the white straight line.
The potential barriers among the grains can be approximated by 
$V_{gb}(p,T)$ and confines the holes, generating local or
intragrain single-particle bound states.
The signature of these single-particle bound states will 
appear in the local density of states together with the
superconducting gap from the two-body attraction. 
Second, and more important, {\it to minimize the local free
energy, the holes move to equilibrium positions in a similar
fashion as if they attract themselves}. This is schematically illustrated
in the low inset of Fig.(\ref{EV6200}) where a) represents
a homogeneous system with $p=0.25$ (one hole at each four sites)
and b) the motion toward clusters formation of low ($p_i=0$) and
high densities ($p_i=0.50$), according to the CH diffusion 
process. Clearly, this movement of holes can 
be regarded as originated  from {\it an effective two-body
attraction}, like the spin-spin exchange interaction, 
that arises from the Pauli principle and the electrons 
wave functions. We do not know the exact strength of
this potential but it should scale with the potential
barrier $V_{gb}(p,T)$ between the two phases. We discuss below
that this effective potential can give rise to 
superconducting pairs.

\section{The Self-consistent Calculations}

The calculated  $u(i,T)$ (or $p({\bf x_i},T)$) on square lattices like the 
$42 \times 42$ shown in Fig.(\ref{MapVu}) is used as
the {\it initial input} and {\it it is  maintained fixed} throughout the
self-consistent Bogoliubov-deGennes (BdG) calculations. 
The pairing potential $V_{gb}(p,T)$ is adjusted to 
match the average local density of states
(LDOS) measured by low temperature STM  on $0.11 \le p \le 0.19$
Bi2212 compounds\cite{McElroy}, namely,
$V_{gb}(p,T=0)=(0.7-2.36\times p)$ (eV). Starting with an extended Hubbard
Hamiltonian, the BdG equations 
are\cite{Mello09,Mello04,DDias08,DDias07,Mello08,Caixa07}.
\begin{equation}
\begin{pmatrix} K         &      \Delta  \cr\cr
           \Delta^*    &       -K^*
\end{pmatrix}
\begin{pmatrix} u_n({\bf x_i})      \cr\cr
                v_n({\bf x_i})
\end{pmatrix}=E_n
\begin{pmatrix} u_n({\bf x_i})       \cr\cr 
                 v_n({\bf x_i})
\end{pmatrix}
\label{matrix}
\end{equation}
These equations, defined in detail in Refs.\cite{Mello04,DDias08,DDias07,Mello08,Caixa07},
are solved self-consistently. $u_n$, $v_n$ and $E_n \ge 0$ are 
respectively the eigenvectors and eigenvalues. As mentioned,
the hole clustering process is mainly along the $Cu-O$ bonds,
and with the Coulomb repulsion, it favours the d-wave pairing. Thus, the
d-wave pairing amplitudes are given by
\begin{eqnarray}
\Delta_{d}({\bf x}_i)&=&-{V_{gb}\over 2}\sum_n[u_n({\bf x}_i)v_n^*({\bf x}_i+{\bf \delta})
+v_n^*({\bf x}_i)u_n({\bf x}_i  \nonumber \\
&&+{\bf \delta})]\tanh{E_n\over 2k_BT} ,
\label{DeltaV}
\end{eqnarray}
and the inhomogeneous hole density is given by
\begin{eqnarray}
p({\bf x}_i)=1-2\sum_n[|u_n({\bf x}_i)|^2f_n+|v_n({\bf x}_i)|^2(1-f_n)],
\label{density}
\end{eqnarray}
where $f_n$ is the Fermi function. We stop the self-consistent calculations
only when all $p({\bf x}_i)$ converges to the CH density map shown 
in Fig.(\ref{MapVu}).

Due to the $Cu$ d-orbitals and the strong on site Coulomb
repulsion, we calculate only the intra-grain d-wave superconducting gap 
$\Delta_d(i,T)$. 
The effect of the temperature in the potential is taken
into account by $V_{gb}(T)\sim (1-(T/T_{PS})^{1.5}$, 
as demonstrated by  CH\cite{CH} near $T_{PS}$, and
consequently, $\Delta_d(i,T)\rightarrow 0$ at a single
temperature $T^*(p)$. The fact that all gaps vanishes at
the same temperature although the values of $\Delta_d(i,T=0)$
can be very different is a consequence of the self-consistent
mean field approach.

As the top inset of Fig.(\ref{EV6200}) shows, the 
different local densities regions are bounded by the potential
barriers $\approx V_{gb}$. As the temperature goes down, the
barriers increases forming an metal-insulator-metal junction. 
Consequently, the electronic structure
in the $CuO-2$ planes becomes similar to that of a granular 
superconductor\cite{Merchant}. In this way the
superconducting transition occurs in two steps: first by the
appearing of intra-grain superconductivity in the grains and 
than by Josephson coupling with phase locking among all the
hole-rich and hole-poor regions at a lower temperature. 

\begin{figure}[ht]
\begin{center}
  \begin{minipage}[b]{.1\textwidth}
    \begin{center}
     \centerline{\includegraphics[width=7.0cm]{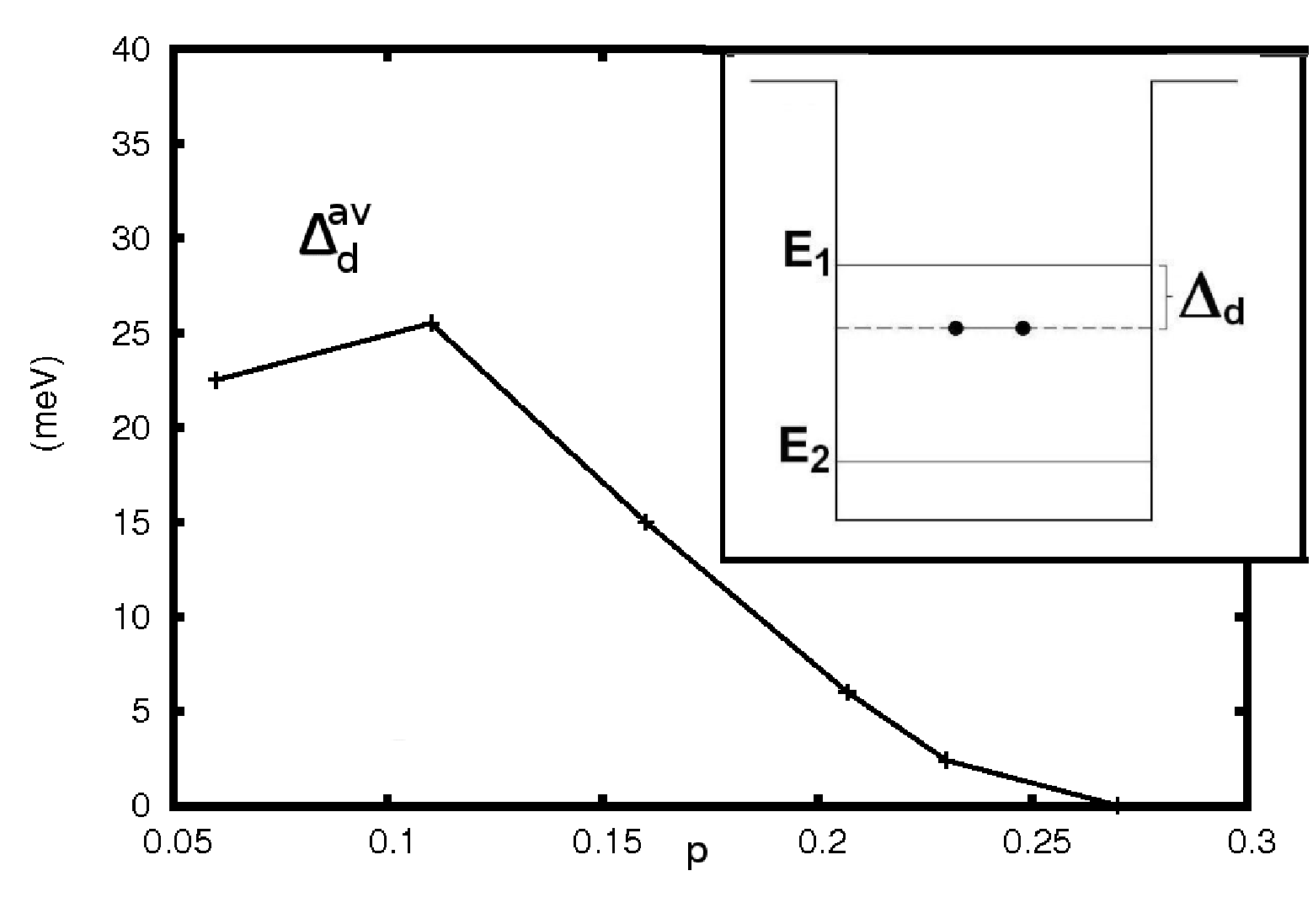}}
    \centerline{\includegraphics[width=7.0cm]{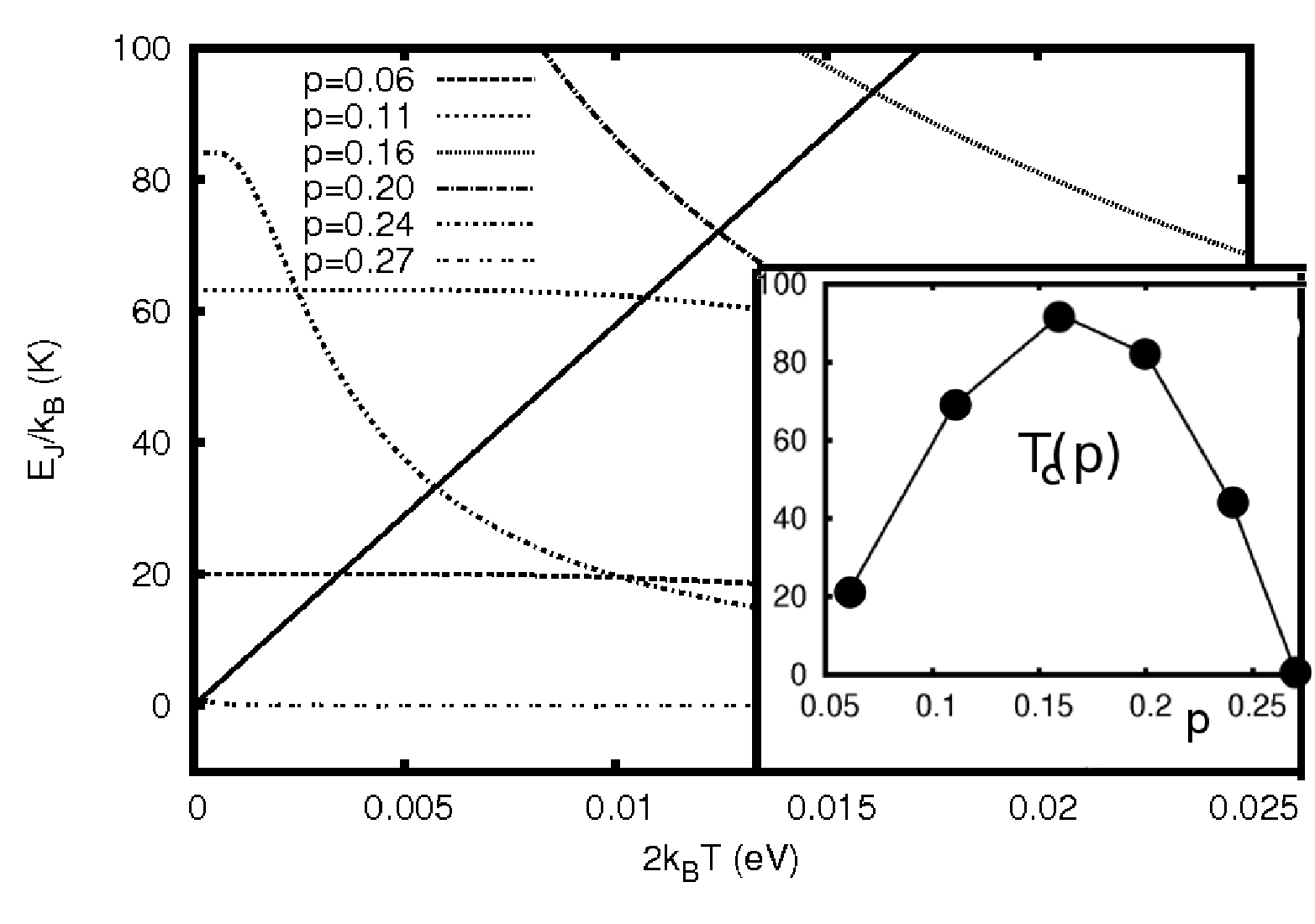}}
    \end{center}
  \end{minipage}
\caption{Top panel, the average $\Delta_d(p)$, and in the
inset, the schematic single particle and superconducting energy levels 
present in each nanoscopic grain. In the low panel, the thermal energy
$k_BT$ and the Josephson coupling among superconducting grains $E_J(p,T)$
for some selected doping values as function of T. The intersections
give $T_c(p)$, as plotted in the inset.  }
\label{EJTc}
 \end{center}
\end{figure}

By using the theory of granular superconductors\cite{AB} to these
electronic grains and the calculated average superconducting
amplitudes $\Delta^{av}_d(T,p)\equiv \sum_i^N \Delta_d(T,i,p)/N$, where N is
the total number of sites, we can estimate the values of $T_c(p)$.
\begin{eqnarray} E_J(p,T) = \frac{\pi h\Delta^{av}_d(T,p)}{4 e^2 R_n}
tanh(\frac{\Delta^{av}_d(T,p)}{2K_BT_c}).
\label{EJ} 
\end{eqnarray} 
Where
$\Delta^{av}_d(T,p)$ is the average of the local superconducting gaps
$\Delta_d(i,p,T)$ on a $N \times N$ ($N=28$, $36$ and $42$) square
lattice which is plotted in the top panel of Fig.(\ref{EJTc}). The inset
shows the schematically single particle levels at a shallow 
puddle whose  walls are proportional to $V_{gb}$. $R_n$ is 
the normal resistance of a given compound, which
is proportional to the planar resistivity $\rho_{ab}$ 
measurements\cite{Takagi} on the
$La_{2-p}Sr_pCuO_2$ series. 

In the low panel of Fig.(\ref{EJTc}),
the Josephson coupling $E_J(p,T)$ is plotted together with the
thermal energy $k_BT$ whose intersection yields the critical temperature
$T_c(p)$, as shown in the inset. The fact that $T_c(p)$ has the
well known dome shape is due to the different behavior of $\Delta^{av}_d(T,p)$
that decreases with $p$ (Fig.(\ref{EJTc})) and $1/R_n$ that
increases with $p$. The
values are in reasonable agreement with the
Bi2212 $T_c(p)$, as expected, since $V_{gb}$ was chosen to 
match the Bi2212 low temperature LDOS gaps\cite{McElroy}.
Since, in general
$T^*(p))>T_c(p)$, this approach {\it
provides an interpretation to the superconducting amplitude
and the  measured quasiparticles dispersion in the normal
phase}\cite{Campuzano,Chatterjee}.

\section{The LDOS and superconducting gap}
In the BdG
approach, the symmetric local density of states (LDOS) is
proportional to the spectral function\cite{Gygi} and may be written
as \begin{eqnarray}
 N_i(T,V_{gb},eV)&=&\sum_n[|u_n({\bf
x}_i)|^2 + |v_n({\bf x}_i)|^2]\times  \nonumber \\
&& [f_n^{'}(eV-E_n)+f_n^{'}(eV+E_n)]. \label{LDOS}
\end{eqnarray}
The prime is the derivative with respect to the argument. 
$u_n, v_n$ and $E_n$ are respectively the eigenvectors and
eigenvalues of the BdG matrix
equation\cite{Mello04,DDias08}, $f_n$ is the
Fermi function and $V$ is the applied voltage. 
$N_i(T,V_{gb},eV)\equiv LDOS(V_{gb})$ is proportional to the
tunneling conductance $dI/dV$, and we probe the effects of the 
inhomogeneous potential by examining the ratio $LDOS(V_{gb}\ne
0)/LDOS(V_{gb}=0)$. This LDOS ratio yields well-defined peaks 
and converges to the unity at large bias,
similar LDOS ratios calculated from STM 
measurements\cite{Gomes,Pasupathy}.

\begin{figure}[ht]
\begin{center}
     \centerline{\includegraphics[width=6.0cm,angle=-90]{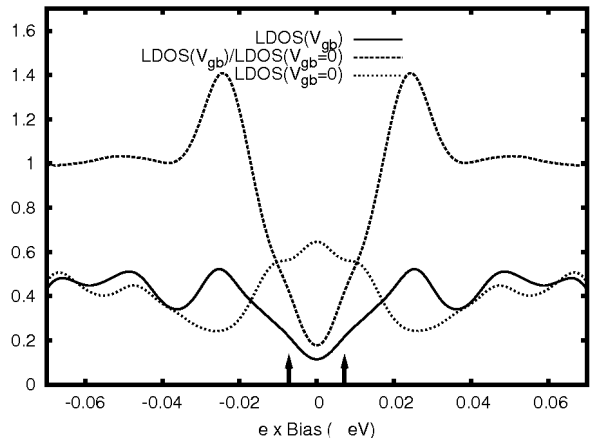}}
\caption{ The LDOS on an overdoped $p=0.20$ sample 
for $V_{gb}=0.240$eV and for $V_{gb}=0$ and their
ratio. The arrows show the values of  the superconducting gap
$\Delta_d=\pm 7.2meV$, much smaller than the LDOS gap ($\approx 24$meV),
which is normally measured by ARPES and STM experiments.  } 
\label{LDOS36n20T40R} 
\end{center}
\end{figure}

These features are illustrated in Fig.(\ref{LDOS36n20T40R}) 
for a representative overdoped $p=0.20$ compound near a grain boundary at
a representative  average doping hole point. At $T=40$K, 
$V_{gb}=0.240$eV  and for large applied potential difference 
($V>0.1$eV) both LDOS, $LDOS(V_{gb}\ne 0)$ and
$LDOS(V_{gb}=0)$ converge to the same values. 
An important result of our calculations is that
the LDOS yield peaks at larger bias than the values of 
the superconducting amplitude $\Delta_d(i,p,T)$. In Fig.(\ref{LDOS36n20T40R})
we can see the LDOS gap  $\Delta_{PG}=24$meV (Eq.(\ref{LDOS}) and the much
smaller superconducting gap $\Delta_d=7.2$meV (Eq.(\ref{DeltaV})) that
is almost undiscernable and therefore it is marked by arrows.
We attribute $\Delta_{PG}$ to the single-particle bound states in
the grains. In our calculations we can see that hole-poor 
produces larger and not well defined peaks than those at hole-rich regions
as noticed by McElroy et al\cite{McElroy}. The physical explanation
is that, if the number of holes in a grain is large (hole-rich grain) 
they occupy higher single-particle levels in the local potential
wells and can be more easily removed by the STM tip.

\begin{figure}[ht]
\begin{center}
     \centerline{\includegraphics[width=8.0cm]{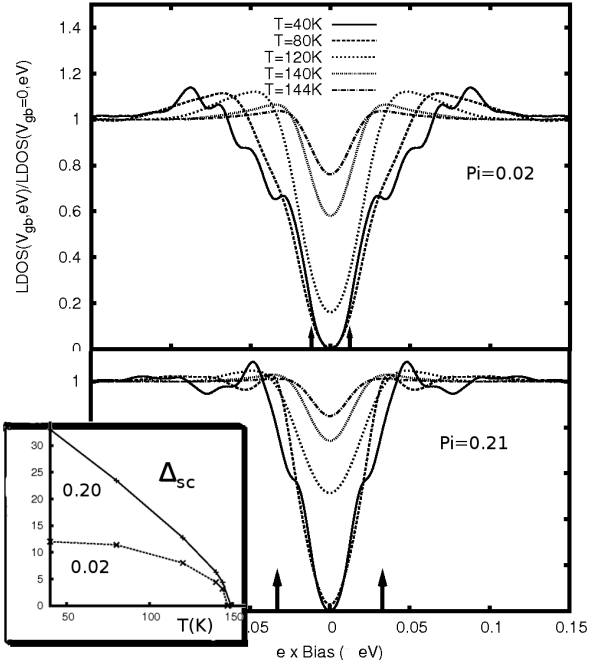}}
\caption{The $p=0.11$ LDOS with $V_{gb}=0.44$eV. 
Top panel, $\Delta_{PG}(T)$ at a hole-poor puddle
($p_i=0.021$). 
$\Delta_{PG}(T=40K)\approx 80$meV. The superconducting 
gaps $\Delta_d(T=40K)=12$meV are marked by arrows.  
Below, the set of LDOS curves at a hole-rich
grain, ($p_i\approx 0.20$) with  $\Delta_{PG}(T=40K)\approx 50$meV
and with $\Delta_d(T=0)=33meV$. 
In the inset $\Delta_d(i,T) \times T$ for each case. Both
$\Delta_{PG}(i)$ and $\Delta_{d}(i)$ vanish near $T^*=147$K.} 
\label{LDOSn11} 
\end{center}
\end{figure}

To study the 
pseudogap phase we perform calculations on an 
underdoped compound with $p=0.11$ and $T_c\approx65$K 
according Fig.(\ref{EJTc}).  In Fig.(\ref{LDOSn11}) 
we  show the results at two representative locations in 
the hole-rich and hole-poor phases. The
hole-poor region with $p_i=0.021$ has a very large LDOS
of $\Delta_{PG}\approx 80$K and $\Delta_d=12$meV, as shown in the inset.
At a hole-rich location  ($p(i)\approx 0.23$)  
the  LDOS gap is given by $\Delta_{PG}\approx 50$meV with
well-defined peaks at low temperature, 
and the superconducting gap is $\Delta_d(T=0)=33$meV.
Perhaps due to the mean-field type calculations, all the
gaps vanish near $T\approx 147$K, what can be assigned 
the pseudogap temperature $T^*(p=0.11)$. Likewise
Fig.(\ref{LDOS36n20T40R}) the low temperature
superconducting gap $\Delta_d$ produces small anomalies
that are marked by arrows, as it was already noticed by our
previous work\cite{Mello09} and detected experimentally  by
Kato et al\cite{Kato,Kato2}.

\begin{figure}[ht]
     \centerline{\includegraphics[width=6.0cm,angle=-90]{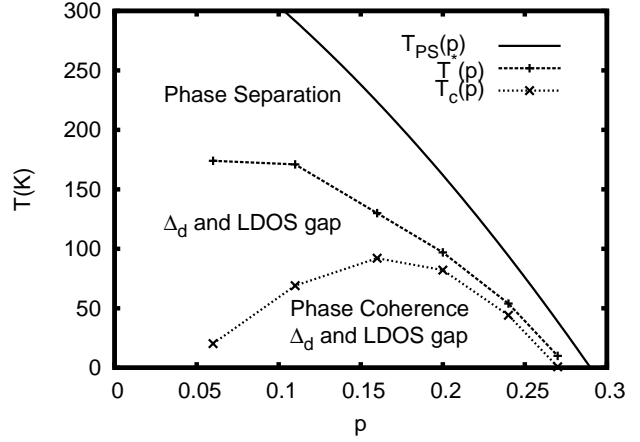}}
\caption{The calculated phase diagram of cuprate superconductors as derived from
the EPS transition $T_{PS}(p)$ and the formation of tiny grains. 
The onset of single particle bound states and superconducting amplitude 
formation occurs at $T^*(p)$. The dome shape $T_c(p)$ curve is due to 
Josephson coupling among the grains. These three lines are in agreement
with the experimental results. }
\label{PhaseDiag} 
\end{figure}

In this scenario the pesudogap phase is formed with the intragrain
superconducting amplitude and the single-particle bound states. 
The superconducting phase also contains both 
the local single-particle gap $\Delta_{PG}(i,p,T)$
and the local superconducting $\Delta_{d}(i,p,T)$ but has
phase coherent through the Josephson coupling among the grains. At the 
overdoped region  $T^*(p)$ approaches $T_c(p)$ while their difference
increases in the underdoped region. These two curves are
shown in the derived phase diagram of Fig.(\ref{PhaseDiag}). We have
also plotted the EPS line $T_{PS}(p)$ which opens
all the phase separation process and it is close the observed
anomaly called upper-pseudogap Timusk and Statt\cite{TS}.

\section{Conclusions}

The main idea of this letter is the two-body electronic attractive potential 
and granular structure derived from the EPS transition.
We show that an effective hole-hole attraction appears
from the Ginzburg-Landau free energy and the Cahn-Hilliard 
simulations which reveals also a microscopic granular behavior 
with single-particle bound states (as seen in Fig.(2)).
With the values of $V_{gb}(p)$ that reproduced
the low temperature average LDOS values of McElroy et al\cite{McElroy}
and the values of $T_{PS}(p)$ from the crossover or upper pseudogap 
line\cite{TS,Tallon,Zawa}, we obtain the superconducting and pseudogap phases for
{\it all doping $p$}. In summary:

{\it i}) The phase separation transition described by the
Cahn-Hilliard equation minimizes the local free energy
generating separated regions of different local densities.\\
{\it ii}) $\Delta_d$  and $\Delta_{PG}$ have completely different
nature but they have the same origin, namely, the inhomogeneous
potential $V_{gb}$. They are both present in the pseudogap and
superconducting phases but distinct strength. Their differences 
were verified in the presence of strong 
applied magnetic fields and also by  their distinct temperature 
behaviour in tunneling experiments\cite{Krasnov}.\\
{\it iii}) The LDOS  are dominated by the intragrain single-particle bound
states $\Delta_{PG}$ and the superconducting gap $\Delta_d$
produces only a small anomaly at low temperature and low bias 
as recently observed by some STM data\cite{Gomes,Pushp,Kato,Kato2}.\\
{\it iv}) The observed difference in the
LDOS shape, called "coherent" and "zero temperature 
pseudogap"\cite{McElroy}, is due to the differences in the
local single-particle bound levels  
at hole-rich and  hole-poor locations respectively.\\
{\it v}) The potential barriers $V_{gb}$ prevents 
phase coherence and, 
as in a granular superconductor, the resistivity 
transition occurs due to Josephson coupling  among the intragrain 
superconducting regions (Eq.\ref{EJ}). The $T_c(p)$ curve is a 
consequence of the different behavior of $1/R_n(p)$ which increases, and 
$\Delta^{av}_d(p)$ which decreases with $p$.\\

We gratefully acknowledge partial financial aid from Brazilian
agency CNPq. 
%
%


\begin{thebibliography}{0}
%

%
%
%
%
\bibitem{TS} \Name{Timusk T. \and Statt B.}
\REVIEW{Rep. Prog. Phys. } {62} {1999} {61}.
%
\bibitem{Tallon} \Name{ Tallon  J.L.  \and  Loram J.W} 
\REVIEW{Physica C} {349} {2001}{53}.
%
\bibitem{Zawa} \Name{Huefner S. et al}
\REVIEW{Rep. Prog. Phys} {71} {2008} {062501}.
%
\bibitem{Tranquada}\Name{Tranquada J.M. et al} 
\REVIEW{Nature (London)} {375}{1995}{561}.
%
\bibitem{Bianconi}\Name{Bianconi A. et al}
\REVIEW{ Phys. Rev. Lett.} {76}{1996} {3412}.
%
\bibitem{Singer}\Name {Singer P.M., Hunt A.W. \and Imai T.}
\REVIEW{ Phys. Rev. Lett.} { 88} {2002} {47602}.
%
%
\bibitem{Pan}\Name{ S. H. Pan et al}
\REVIEW{Nature} {413} {2001} {282}.
%
\bibitem{McElroy}\Name{ McElroy K. , et al}
\REVIEW{Phys. Rev. Lett.}{94} {2005} {197005}.
%
\bibitem{Gomes} \Name{ Gomes Kenjiro K. et al}
\REVIEW{Nature }{ 447} {2007} {569}.
%
\bibitem{Pasupathy}\Name{ Pasupathy Abhay N. et al}
\REVIEW{Science } {320} {196} {2008}.
%
\bibitem{Kato}\Name{ Takuya Kato et al}
\REVIEW{ J. Phys. Soc. Jpn.} {77} {2008} {054710}.
%
\bibitem{Pushp}\Name{ Aakash Pushp et al}
\REVIEW{Science} {324} {2009} {1689}.
\bibitem{Kato2}\Name{ T. Kato et al}
\REVIEW {J. Supercond. Nov. Magn.} {23}{2010} {771}.
%
%
\bibitem{Keren}\Name {Rinat Ofer, and Amit Keren}
\REVIEW {Phys. Rev.} {\bf B80} {2009} {224521}.
%
\bibitem{Bianconi2}\Name {Fratini M., et al}
\REVIEW{Nature} {456} {2010} {841}.
%
\bibitem{LeTacon}\Name{ M. Le Tacon, et al} 
\REVIEW {Nature Phys.} {2} {2006} {537}.
%
\bibitem{Shen}\Name {W. S. Lee, et al}
\REVIEW {Nature} {450} {2007} {81}.
%
\bibitem{Mad}\Name {J. H. Ma, et al} 
\REVIEW {Phys. Rev. Lett.} {101} {2008} {207002}.
%
\bibitem{Campuzano}\Name{Kanigel A. et al}
\REVIEW{
Phys. Rev. Lett.}{101}{2008}{ 137002}.
%
\bibitem{Chatterjee} \Name{U. Chatterjee, et al}
\REVIEW {Nature Phys.} {6} {2010} {99}.
%
%
\bibitem{Mello09}\Name{de Mello E.V.L, Passos C.A.C \and Kasal R.B.}
\REVIEW{J. Phys.: Condens. Matter} {21} {2009} {235701}.
%
\bibitem{Tranquada2} \Name{S. Wakimoto et al}
\REVIEW {Phys. Rev. Lett.} {98} {2007} {247003}.
%
%
%
%
\bibitem{Bray} \Name{Bray A.J. }
\REVIEW{Adv. Phys.}{ 43}{1994}{347}.
%
\bibitem{Muller}\Name{Sigmund E. \and Muller K.A (Eds.}
\Book {Phase Separation in Cuprate Superconductors}
\Publ{Springer-Verlag, Berlin} \Year{1993}.
%
\bibitem{CH} \Name{Cahn  J.W. \and Hilliard J.E. } 
\REVIEW{J. Chem. Phys} {28}{1958} {258}.
%
\bibitem{Otton}\Name{de Mello E.V.L. \and Silveira Filho  Otton T. }
\REVIEW{Physica A}{347}{2005} {429}.
%
\bibitem{Mello04} \Name{ de Mello E.V.L. \and  Caixeiro E.S.}
\REVIEW{ Phys. Rev. B}{70}{2004}{224517}.
%
%
%
\bibitem{DDias08} \Name{ Dias D. N. et al}
\REVIEW{Physica C} {468}{2008} {480}.
%
%
\bibitem{DDias07}\Name{  de Mello E. V. L. \and Dias  D. N.}
\REVIEW{J. Phys.C.M. }{ 19}{2007} {086218}.
%
\bibitem{Mello08}\Name{de Mello E.V.L. et al} 
\REVIEW{Physica B} {404} {2009}{3119}.
%
\bibitem{Caixa07}\Name{Caixeiro E.S, de Mello E.V.L. \and Troper A,}
\REVIEW{ Physica C}{459} {2007} {37}.
%
%
\bibitem{Merchant} \Name{Merchant L. et al}
\REVIEW{Phys. Rev.B}{63} {2001}{134508}.
%
%
%
%
%
%
%
\bibitem{AB} \Name{Ambeogakar V. \and Baratoff A.}
\REVIEW{ Phys. Rev. Lett. }{10}{1963}{486}.
%
\bibitem{Takagi}\Name{Takagi H. et al}
\REVIEW{Phys. Rev. Lett. }{69} {1992}{2975}.
%
\bibitem{Gygi}\Name{Gygi Fran\c{c}ois \and Schl\"uter Michel}
\REVIEW {Phys. Rev. B} {43} {1991} {7609}.
%
%
\bibitem{Krasnov}\Name{Krasnov V.M. et al}
\REVIEW {Phys. Rev. Lett.} {86} {2001} {2657}.
\end{thebibliography}
\end{document}